\documentclass{article}
\usepackage{hiph-artx}
\newcommand{\be}{\begin{equation}}
\newcommand{\ee}{\end{equation}}
\newcommand{\ba}{\begin{eqnarray}}
\newcommand{\ea}{\end{eqnarray}}

\volnumber{27} \issuenumber{1} \edyear{2006}                             
\frompage{459} \topage{468}                                              
\recrevdate{15 September 2006}                                           
\def\rightmark{Kaon and Pion Ratio Probes of Jet Quenching}
\def\leftmark{P. L\'evai et al.}

\title{Kaon and Pion Ratio Probes of Jet Quenching \\ in Nuclear Collisions }
\authors{
{P. L\'evai$^{1,2,3}$, G. Papp$^{2,4}$, G. Fai$^{2}$,
and M. Gyulassy$^{3}$
\index{L\'evai, P.} 
\index{Papp, G.} 
\index{Fai, G.}
\index{Gyulassy, M.}
}\\[2.812mm]
{\normalsize
\hspace*{-8pt}$^1$ KFKI RMKI 
Research Institute for Nuclear and Particle Physics, \\
49 PO Box, Budapest, 1525, Hungary \\
\hspace*{-8pt}$^4$ Department of Physics, Kent State University, 
   Kent, 44242, OH, USA \\
\hspace*{-8pt}$^2$ Department of Physics, Columbia University,  \\
        538 W 120th Street, New York, NY 10027, USA\\
\hspace*{-8pt}$^3$ Department of Theoretical Physics,  
  E\"otv\"os University,\\\ \ \ P\'azm\'any P. 1/A, Budapest, 1117, Hungary 
}}

\abstract{
Non-abelian energy loss in quark gluon plasmas
 is shown to lead to novel hadron ratio suppression patterns
 in ultrarelativistic nuclear collisions.
Here we investigate pion and kaon production
in $pp$ and $AA$ collisions in a perturbative QCD frame,
suppression pattern and hadron ratios.
The $K^-/K^+$ and $K^+/\pi^+$ ratios
are found to be  most sensitive to the opacity (density) of the plasma. 
Experimental data indicate that the fragmentation
dominated pQCD region will be reached only
at higher $p_T$; in an intermediate $p_T$ region other particle 
production mechanisms dominate the $K/\pi$ ratios.}


\begin{document}

\maketitle
\setcounter{page}{1}

Energy loss of high energy quark and gluon jets penetrating  dense
matter produced in ultrarelativistic heavy ion collisions
leads to jet quenching and thus  probes the quark-gluon plasma formed 
in those reactions~\cite{gptw,mgxw92,bdms,bdms8,zahar,urs00,glv2,glv2b}.
The non-abelian radiative energy loss suppresses the hadron production
yield in the momentum range $2-3 \ {\rm GeV/c} < p_T < 15-20 \ {\rm GeV/c}$ 
at RHIC energies~\cite{PHENIXAuAu,STARAuAu}.  
Since quark and gluon jets suffer different 
energy losses proportional to their color Casimir factors $(4/3,3)$,
the energy loss effect could be investigated through
the jet fragmentation pattern into hadrons with different flavor content. 
Here we present a perturbative QCD based calculation of different 
hadron ratios in $p+p$ collisions~\cite{PLF00,YZ02}, 
then extend our calculation for
heavy ion collisions and include jet energy loss.

Let us summarize our basic knowledge about non-abelian energy loss in
hot dense matter.
First estimates~\cite{gptw,mgxw92} suggested a linear dependence
on the plasma thickness, $L$, namely
$\Delta E \approx 1-2 {\rm \ GeV}(L/{\rm fm})$,
as in abelian electrodynamics.
In BDMS~\cite{bdms,bdms8}, however, non-abelian (radiated gluon final state
interaction) effects were shown to lead to  a quadratic
dependence on $L$ with a larger magnitude of $\Delta E$.
Similar results have been obtained from light-cone path integral
formalism~\cite{zahar,urs00}.


In the GLV formalism, applying opacity series (see Refs.~\cite{glv2,glv2b}),
finite kinematic constraints were found to reduce greatly the energy 
loss at moderate jet energies. On the other hand,
the quadratic dependence on $L$ has been recovered in wide energy region.
As we show below, the kinematically
suppressed GLV energy loss turns out to depend
 approximately linearly on the jet energy, $E$. 
This linear dependence, $\Delta E\propto E$, unfortunately leads
to a very weak $p_T$-dependent
suppression of the transverse momentum distributions for
the kinematic range accessible experimentally at RHIC.
Our focus is to investigate whether
the $p_T$ dependence of the particle ratios 
may help to pin down  the jet quenching mechanisms.
Our primary candidates are the  measurable $K^+/\pi^+$ and $K^-/K^+$
ratios. 

Non-abelian energy loss in pQCD has been  calculated analytically
in two limits. In the ``thick plasma'' limit,
the mean number of jet scatterings,
$\bar{n}=L/\lambda$, is assumed to be much greater than one.
For asymptotic jet energies the eikonal
approximation  applies and the resummed
energy loss (ignoring kinematic constraints) reduces to the
following simple form~\cite{bdms,bdms8,zahar,urs00}:
\begin{equation}
\Delta E_{BDMS}=\frac{C_{R}\alpha_s}{4}\,
\frac{L^2\mu^2}{\lambda_g} \,\tilde{v} \;\;,
\label{de1}
\end{equation}
where $C_R$ is the color Casimir of the jet ($=N_c$ for gluons),
and $\mu^2/\lambda_g\propto \alpha_s^2\rho$ is a  transport
coefficient of the medium proportional to the parton density, $\rho$.
The factor, $\tilde{v}\sim 1-3$  depends
logarithmically on $L$ and the color Debye screening scale, $\mu$.
It is the radiated gluon mean free path, $\lambda_g$, that enters above.

In the ``thin plasma''  approximation~\cite{glv2,glv2b},
the opacity expansion was applied 
and in the first order the following expression was derived 
for the energy loss:
\begin{eqnarray}
\Delta E_{GLV}^{(1)}&=& \frac{2 C_R \alpha_s}{\pi} 
\frac{E L}{\lambda_g} \, \int_0^1 dx
\int_0^{k_{max}^2} \frac{d {\bf k}^2_\perp}{{\bf k}^2_\perp}
\nonumber \\[.5ex] 
&& \hspace{-0.5in}  \int_0^{q_{\max}^2} 
\frac{ d^2{\bf q}_{\perp} \, \mu_{eff}^2 }{\pi 
({\bf q}_{\perp}^2 + \mu^2)^2 } \cdot 
\frac{ 2\,{\bf k}_\perp \cdot {\bf q}_{\perp}
  ({\bf k} - {\bf q})_\perp^2  L^2}
{16x^2E^2 \ + \ ({\bf k} - {\bf q})_\perp^4  L^2 } \;\;.          
\label{dnx1}
\end{eqnarray}

\noindent
Here the opacity factor $L/\lambda_g$
is the average number of final state interactions
that the radiated gluons suffer  in the plasma.
The upper transverse kinematic limit is
\begin{equation}
\quad {\bf k}^2_{\max}=\min\, [4E^2x^2,4E^2x(1-x)]\;,
\label{klimits}
\end{equation}
and the upper kinematic bound on the momentum transfer is
 $q^2_{\rm max}= s/4 \simeq 3 E \mu$. 

\newpage
\noindent
Furthermore, 
$\mu_{eff}^2/\mu^2=1+\mu^2/q_{\max}^2$. 
For SPS and RHIC energies,
these finite limits cannot be ignored~\cite{glv2,glv2b}.
The integral  averages over a screened Yukawa interaction with scale $\mu$.
The integrand is for an  exponential density profile,
$\rho\propto \exp(-2z/L)$, with the same mean thickness, $L/2$,
 as a uniform slab of plasma 
of width $L$.

It was shown in Refs.~\cite{glv2,glv2b}
that in the asymptotic $E\rightarrow\infty$
limit, the first-order expression (\ref{dnx1})
reduces to the BDMS result \cite{bdms,bdms8}
up to a logarithmic factor $\log(E/\mu)$. 
Numerical solutions revealed that second and third order 
in opacity corrections to eq. (\ref{dnx1}) remain  small 
$(<\%20)$ in the kinematic range of interest.

For applications to 
RHIC energies ($\sqrt{s}=130-200$~AGeV), we can find
opacity values of $L/\lambda= 3-4$~\cite{BGG06}, which are connected
to gluon rapidity densities of
$dN_g/dy = 1000-1200$~\cite{Vitev05}.  These values are consistent
with the HIJING estimates on gluon 
rapidity densities~\cite{HIJ,Last99} and the measured
charged hadron rapidity densities ($\sim 550-650)$~\cite{PHOBOS00,PHENIX01}.
We illustrate here the jet quenching effects for a generic plasma
with an average
screening
scale $\mu=0.5$~GeV, $\alpha_s=0.3$, and an average
gluon mean free path  $\lambda_g=1$~fm.
The numerical results for the
first order energy loss $\Delta E$, taking into account the finite kinematic
bounds are displayed
in Fig. 1 for different opacities  ${\bar n} = L/\lambda_g=1-4$.
Since $\Delta E\propto C_R$, quark jet energy loss is simply 4/9
of the gluon energy loss shown in Fig. 1.
As noted in Refs.~\cite{glv2,glv2b},
the first order  opacity result
 reproduces the characteristic BDMS quadratic dependence of the energy loss
on $L$.
\vspace{-0.5cm}
\begin{center}
\vspace*{8.90cm}\hspace{-11.5truecm}
\includegraphics{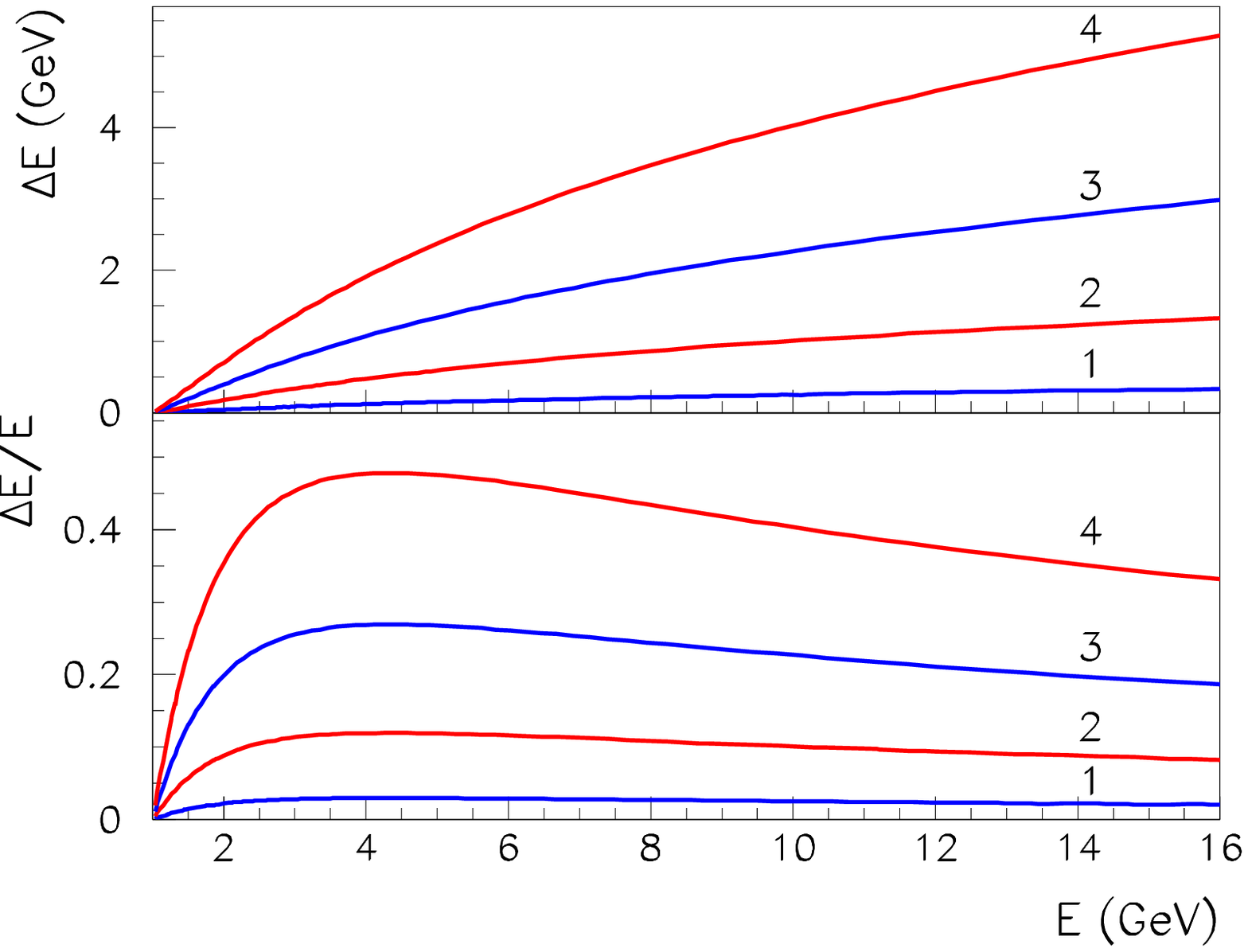}
\vspace{-0.2cm}
\end{center}
\begin{center}
\begin{minipage}[t]{12.6cm}
         { Fig. 1.}
         {\small Absolute ($\Delta E$) 
          and relative ($\Delta E/E$) energy loss of a gluon jet at
          different opacities, ${\bar n} = L/\lambda = 1,2,3,4$,
          as the function of the jet energy.}
\end{minipage}
\end{center}
\vspace{0.3cm}

\newpage

The most surprising feature of Fig. 1 is that
unlike in the asymptotic BDMS case, where $\Delta E/E\propto
1/E$ decreases rapidly with energy, the finite kinematic
bounds in the GLV case give rise to an approximate linear
energy dependence of $\Delta E$ in 
the energy range, $E=2-10$ GeV.
For this plasma, characterized by
the transport coefficient $\mu^2/\lambda_g= 0.25 \; {\rm GeV}^2/{\rm fm}$,
the approximately constant
{\em fractional} energy loss of gluon jets is
\begin{equation}
\Delta E_{GLV}/E \approx \;\left(\frac{\rho}{10\;{\rm fm}^{-3}}\right)
\left(\frac{L}{6\;{\rm fm}}\right)^2
\; \; .
\end{equation}
For the energy range displayed, 
$\Delta E_{BDMS}/E\approx (40 {\rm  GeV}/E)(L/6\;{\rm fm})^2$ for
$\tilde{v}=2$, 
exceeds unity throughout this  energy range.
The GLV expression approaches the asymptotic BDMS result from below
only beyond the range of RHIC experiments.

In order to investigate 
the influence of the GLV  energy-dependent radiative energy loss
on  hadron
production, we apply a perturbative QCD 
(pQCD) based description of
$Au+Au$ collisions, including energy loss prior
to hadronization.
First, we check  that the applied pQCD description 
reproduces data on pion and kaon production 
in $p+p$ collision.  
Our results are based on a leading order (LO) pQCD analysis. 
Detailed discussion of the formalism is
published in Refs.~\cite{PLF00,YZ02}. 
Next to leading order calculations
were performed for pions~\cite{BGG06}, but not for kaons.

Our pQCD calculations 
incorporate the parton transverse momentum 
(``intrinsic $k_T$'') via a Gaussian transverse momentum
distribution $ g(\vec{k}_{T})$
(characterized by
the width $\langle k_T^2 \rangle $) \cite{PLF00,XNWint} as: 
\begin{eqnarray}
\label{fullpipp}
&&E_{h}\frac{d\sigma_h^{pp}}{d^3p} =\!
        \sum_{abcd}\!  
        \int\!\!dx_1 dx_2 d^2k_{T,a}d^2k_{T,b}\
        g(\vec{k}_{T,a}) g(\vec{k}_{T,b}) \times
        \nonumber \\
        && \ \ \ f_{a/p}(x_1,Q^2) f_{b/p}(x_2,Q^2)\
             \frac{d\sigma}{d{\hat t}}
   \frac{D_{h/c}(z_c,{\widehat Q}^2)}{\pi z_c} \,\,\, .
\end{eqnarray}

Here we use LO parton distribution functions 
(PDF) from the MRST98 parameterization \cite{MRS98} and a
LO set of fragmentation functions (FF)~\cite{BKK95}.
The applied scales are $Q=p_{\bf c}/2$ and ${\widehat Q}= p_T/2z_c$.

Utilizing available high transverse-momentum ($2 < p_T < 10$ GeV/c)
$p+p$ data on pion and kaon production at $19 < \sqrt{s} < 63$ GeV 
we can determine the best fitting energy-dependent 
$\langle k_T^2 \rangle$ parameter 
(see Ref.~\cite{PLF00} for further details on $\langle k_T^2 \rangle$).
Fig.~2  shows a general agreement between the data 
and our calculations for
the $\pi^- / \pi^+$ and the $K^-/K^+$ ratios
within the present errors for $p_T \geq 3$ GeV.
We also show  predictions
for  $\sqrt{s}=130 \ (56)$ GeV,
with $\langle k_T^2 \rangle = 3.5 \ (3.0)\  {\rm GeV}^2{\rm /c}^2$ for
$\pi^+$, $\pi^-$ and $K^-$. For $K^+$ we used 
a smaller phenomenological value, $\langle k_T^2 \rangle = 2.5 \ (2.0) \ {\rm
GeV}^2{\rm /c}^2$, as  at lower energies \cite{YZ02}.

Fig. 3 displays the results for $K^-/\pi^-$ and $K^+/\pi^+$.
One can conclude that, while agreement is reasonable for $p_T \geq 4$ GeV, 
there is a systematic  discrepancy at smaller transverse momenta.
High statistics $p+p$ data at RHIC will
be essential to establish an accurate baseline to which $A+A$ must be compared.

\newpage
 
\ \ 
\vspace{0.5cm}
\begin{center}
\vspace*{7.0cm}\hspace{-8.0truecm}
\includegraphics{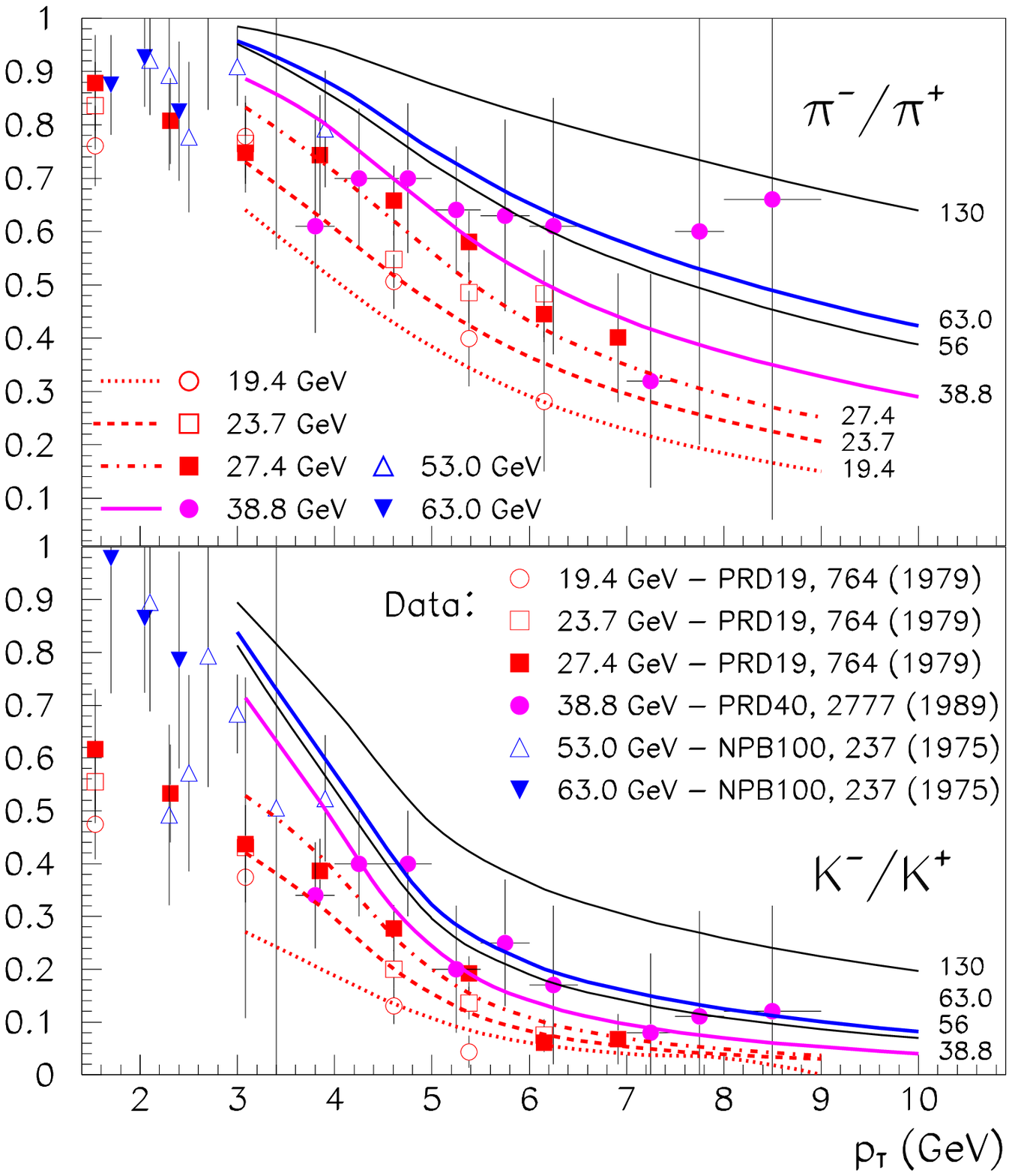}
\vspace{-0.5cm}
\end{center}
\begin{center}
\begin{minipage}[t]{12.6cm}  
      { Fig. 2.} {\small $\pi^-/\pi^+$ and
$K^-/K^+$ ratios in $p+p \rightarrow h+ X$ collisions
in the energy region
$19.4 \leq \sqrt{s} \leq 130$ GeV. Data are from the references
listed in the bottom panel.
} 
\end{minipage}
\end{center}

\vspace{1.5cm}
\begin{center}
\vspace*{7.0cm}\hspace{-8.0truecm}
\includegraphics{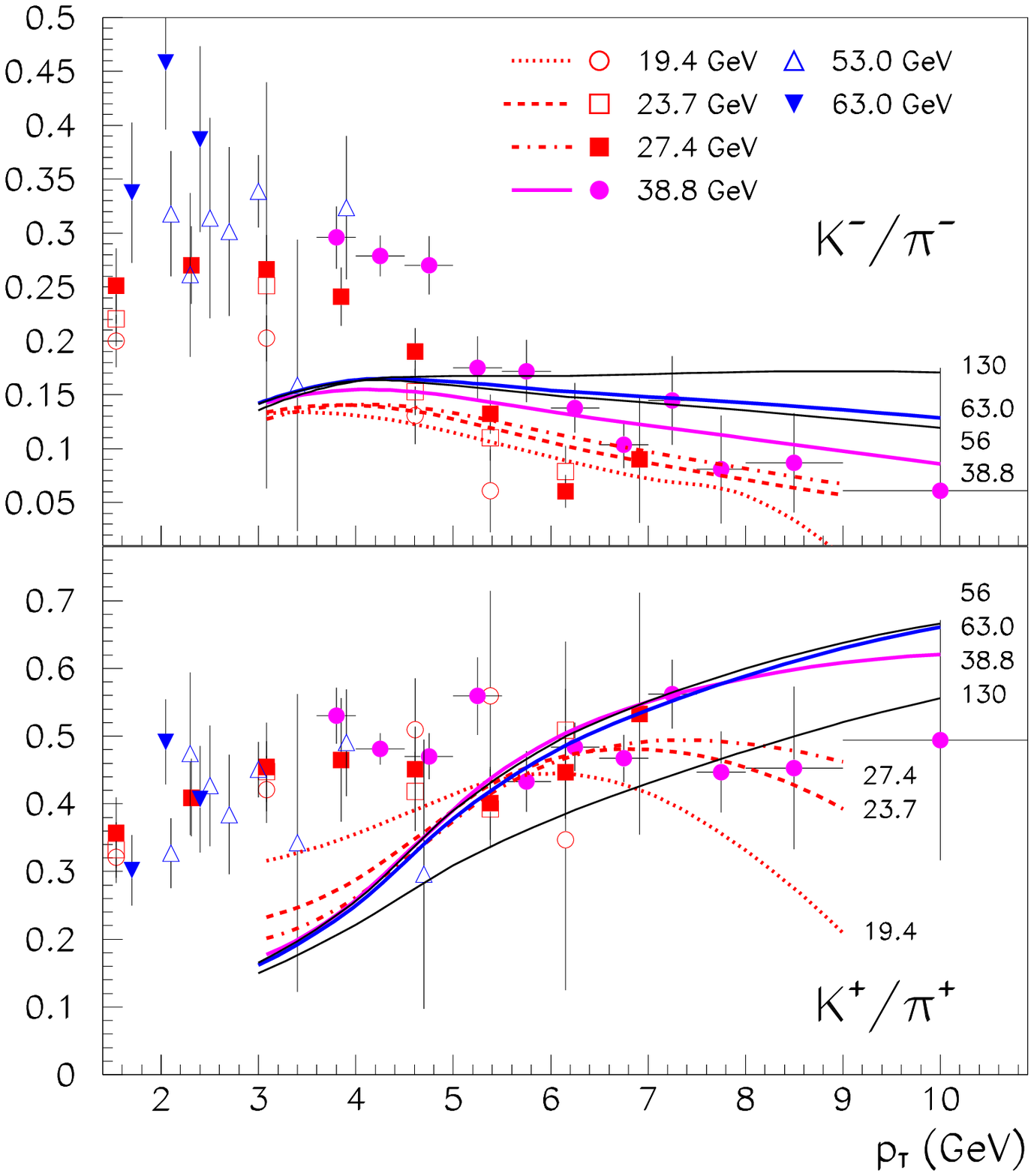}
\vspace{-0.5cm}
\end{center}
\begin{center}
\begin{minipage}[t]{12.6cm}  
      { Fig. 3.} {\small $K^-/\pi^-$ and
$K^+/\pi^+$ ratios in $ p+p \rightarrow h+ X$ collisions
in the energy region
$19.4 \leq \sqrt{s} \leq 130$ GeV. Data are from the references
listed in the bottom panel of Fig. 2.
} 
\end{minipage}
\end{center}

\newpage

Now we turn to
the calculation for $Au+Au$ collision at RHIC energy,
$\sqrt{s} =130$ AGeV. 
As a first approximation,
let us consider 
slab geometry, neglecting radial dependence.
We include the isospin asymmetry and the nuclear
modification (shadowing) into the nuclear PDF.
We consider the average nuclear dependence of the PDF,
and apply a 
scale independent parameterization  with the 
shadowing function $S_{a/A}(x)$ taken
from Ref.~\cite{HIJ}:
\begin{eqnarray}
f_{a/A}(x) &=& S_{a/A}(x)\left[ \frac{Z}{A} f_{a/p}(x) 
+ \left( 1 - \frac{Z}{A} \right) f_{a/n} (x) \right] .
\end{eqnarray}

The value of the
$\langle k_T^2 \rangle$ of the transverse component of the PDF
will be increased by
multiscattering effects in $A+A$ collisions.
Two limiting cases were investigated with 
(i) a large number of rescatterings~\cite{YZ02,XNWint,Papp02} and 
(ii) a small number of rescatterings
(``saturated Cronin effect'') ~\cite{PLF00}.
To clarify the situation,
further study of $p+A$ collisions is necessary.
Here we concentrate
on the influence of jet-quenching on the hadron spectra,
leaving the inclusion of the multiscattering effect for future work.

In this simplified calculation, jet quenching reduces  the energy
of the jet before fragmentation. We concentrate on $y_{cm}=0$,
 where the jet transverse
momentum before fragmentation is shifted by the energy loss,
$p_c^*(L/\lambda) = p_c - \Delta E(E,L)$. This 
shifts the $z_c$ parameter in the integrand
to $z_c^* = z_c /(1-\Delta E/p_c)$.
The applied scale in the FF
is  similarly modified,
${\widehat Q} = p_T/2z_c^*$, while
for the elementary hard reaction the scale
remains $Q=p_c/2$.

With these approximations the invariant cross section of hadron
production in central $A+A$ collision is given by
\begin{eqnarray}
\label{fullaa}
&&E_{h}\frac{d\sigma_h^{AA}}{d^3p} =B_{AA}^{(0)}\!
        \sum_{abcd}\!  
        \int\!\!dx_1 dx_2 d^2k_{T,a}d^2k_{T,b}\
        g(\vec{k}_{T,a}) g(\vec{k}_{T,b})
        \nonumber \\
        && \ \ \ \  f_{a/A}(x_1,Q^2) f_{b/A}(x_2,Q^2)\
             \frac{d\sigma}{d{\hat t}} \frac{z^*_c}{z_c}
   \frac{D_{h/c}(z^*_c,{\widehat Q}^2)}{\pi z_c} \,\,\, ,
\end{eqnarray}
where $B_{AA}^{(0)}= 2 \pi \int_0^{b_{max}} b~db~T_{AA}(b)$ 
with $b_{max}= 3$ fm for central $Au+Au$ collision
and $T_{AA}(b)$ is the thickness function.
The factor $z^*_c/z_c$ appears because of  the in-medium 
modification of the fragmentation function \cite{WaHu97}.
Thus, the invariant cross section 
(\ref{fullaa}) will depend on the average
opacity or collision number, ${\bar n} = L/\lambda_g$.
The calculated spectra for pions and kaons are displayed 
for ${\bar n}=0,2,3,4$ in Fig. 4 (top panels), with
their ratios to the non-quenched spectra (bottom panels).

We note that the GLV energy-dependent
energy loss leads to a rather 
structureless downward shift of the single inclusive
yields of all hadrons because $\Delta E/E$ is approximately
constant. This is in distinction to the estimates
in \cite{XNWint}, where an energy dependent
fractional energy loss,
$\Delta E/E=(0.5 \  {\rm GeV}/E)(L/{\rm fm})$, was used.
 Our new result indicates that
over the entire accessible energy range, the GLV
radiative energy loss 
reduces the $p_T$ distribution
by up to an order of magnitude for pions and somewhat less for kaons.

\newpage

\ \ 
\vspace*{2.1cm}
\begin{center}
\vspace*{8.30cm}\hspace{-11.5truecm}
\includegraphics{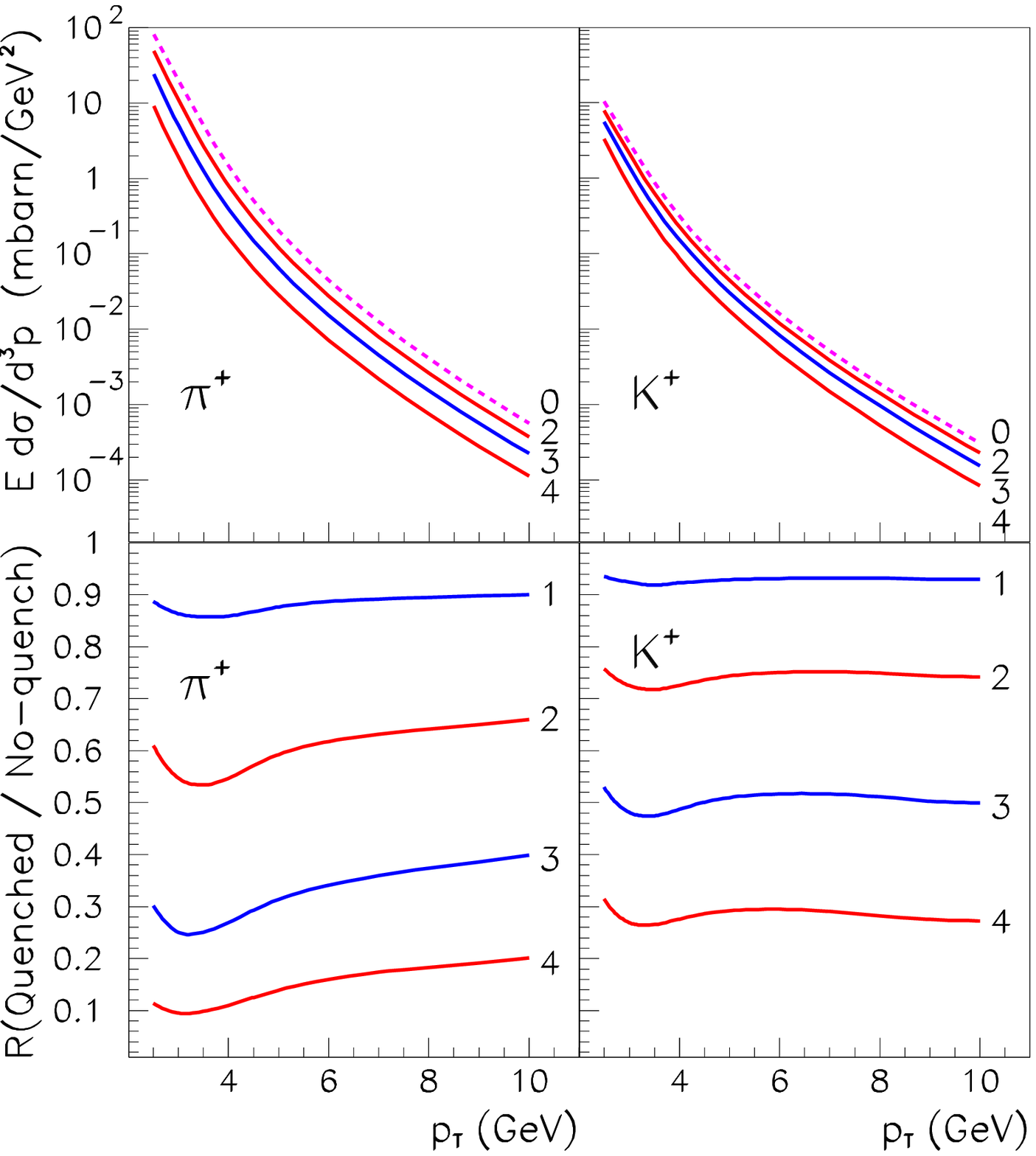}
\vspace{-0.2cm}
\end{center}
\begin{center}
\begin{minipage}[t]{12.6cm}  
      { Fig. 4.} {\small Top: $\pi^+$ and $K^+$
spectra in $Au+Au$ collision at $\sqrt{s}=130 \ AGeV$
 without jet-quenching (dashed line) 
and with ${\bar n}=2,3,4$ (full lines).
Bottom: Ratios of the
spectra with quenching  to the ``non-quenching'' case 
for ${\bar n}=1,2,3,4$.
For $\pi^-$ and $K^-$ one obtains similar spectra and ratios.
} 
\end{minipage}
\end{center}
\vspace{0.3cm}

Fig. 4 indicates the importance of the absolutely normalized hadronic
spectra to study the effect of jet-quenching.
The effect on particle ratios is, however, more interesting, as shown
by Fig.-s 5 and 6. 

While the influence on $\pi^-/\pi^+$ 
is very small, the $K^-/K^+$ ratio is predicted to drop dramatically in the
few GeV domain. The shift of solid $K^-/K^+$ curves to smaller $p_T$
(relative to the dotted curve) can be used to test the nonlinearity
of the energy loss as a function of $L$. It also
provides a constraint on the magnitude
of the $\mu^2/\lambda_g$ transport coefficient.
The most conspicuous quenching effect is limited in the  $K^-/\pi^-$ ratio 
to below $p_T < 4-5$ GeV/c. There is about a factor of 2 
enhancement of the $K^+/\pi^+$ ratio for opacity
${\bar n} = 4$ throughout this range. 

\newpage

\ \
\vspace{0.5cm}
\begin{center}
\vspace*{7.0cm}\hspace{-8.0truecm}
\includegraphics{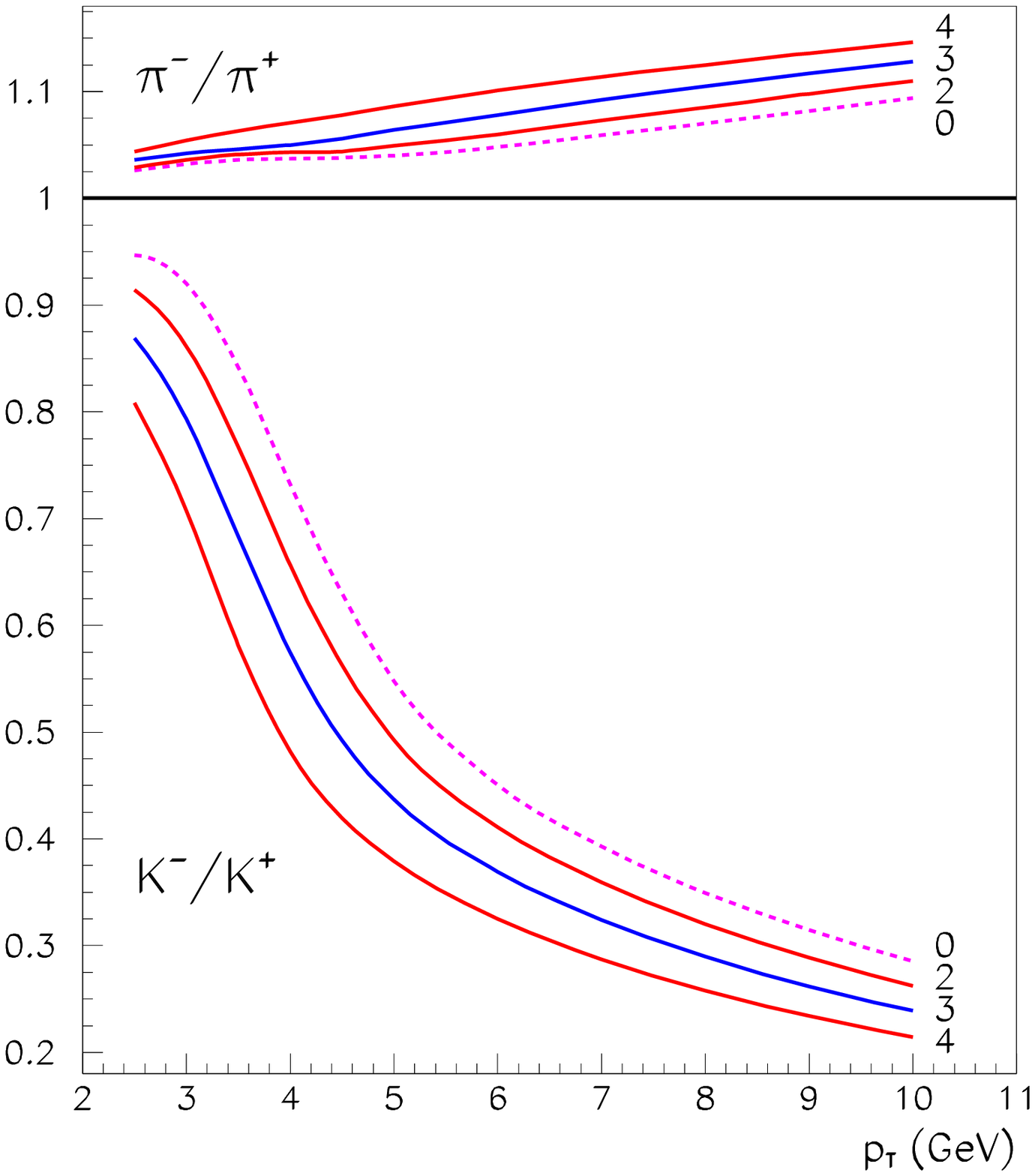}
\vspace{-0.5cm}
\end{center}
\begin{center}
\begin{minipage}[t]{12.6cm}
      { Fig. 5.} {\small $\pi^-/\pi^+$ and
$K^-/K^+$ ratios in $ Au+Au \rightarrow h+ X$
collision at $\sqrt{s}=130 \ AGeV$
without jet-quenching (dotted line)
and with it,  ${\bar n}=2,3,4$.
}
\end{minipage}
\end{center}

\vspace{1.5cm}
\begin{center}
\vspace*{7.0cm}\hspace{-8.0truecm}
\includegraphics{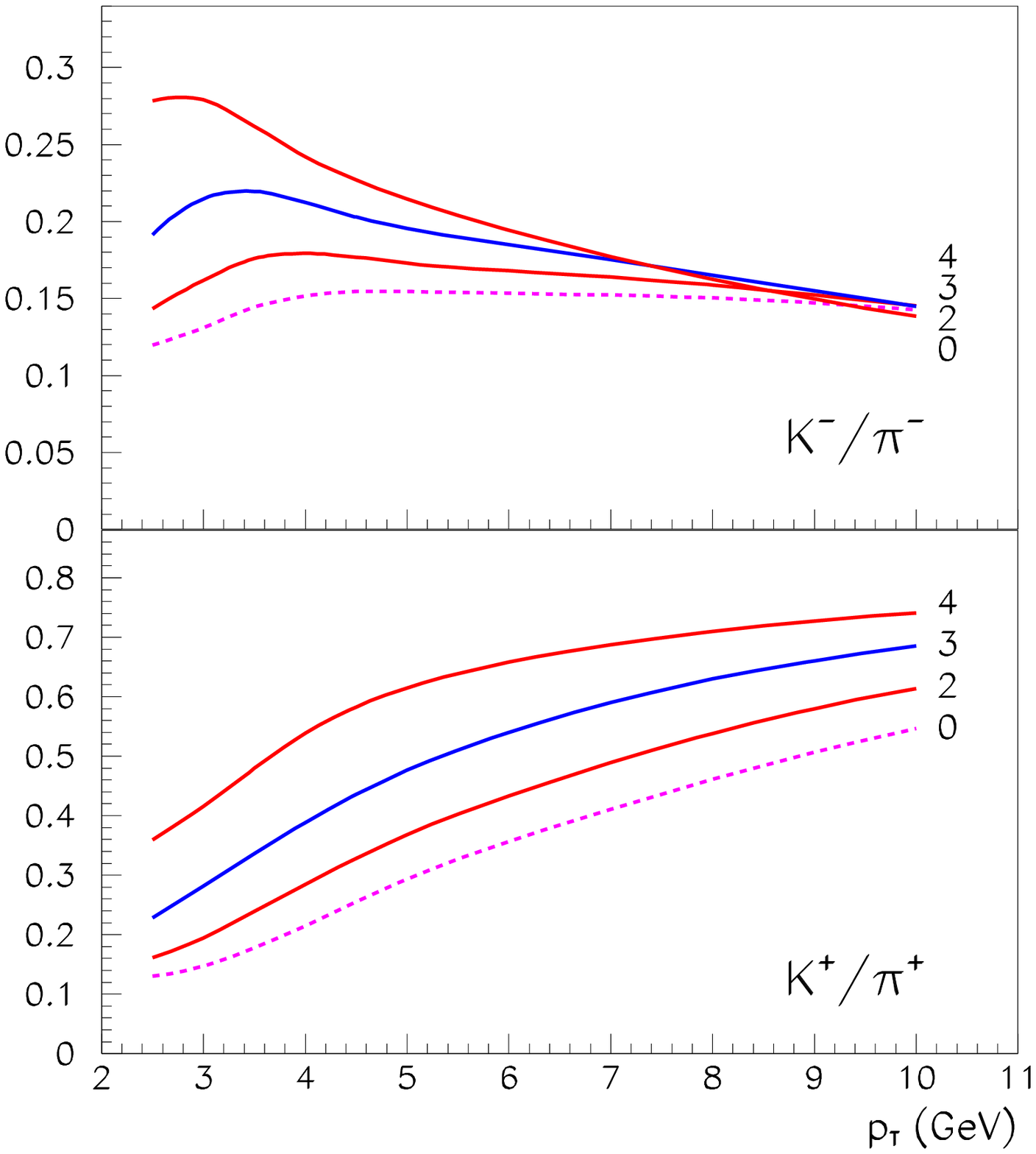}
\vspace{-0.5cm}
\end{center}
\begin{center}
\begin{minipage}[t]{12.6cm}  
      { Fig. 6.} {\small $K^-/\pi^-$ and
$K^+/\pi^+$ ratios in $ Au+Au \rightarrow h+ X$
collision at $\sqrt{s}=130 \ AGeV$
without jet-quenching (dotted line) 
and with it,  ${\bar n}=2,3,4$.
} 
\end{minipage}
\end{center}

\newpage

These patterns arise from the interplay of
the energy-shifted gluon and quark fragmentation functions
into pions and kaons. Since the quark to gluon jet ratio 
at fixed $p_T$ depends sensitively on the beam energy for
$\sqrt{s}=20-200$ AGeV, these patterns are 
expected to shift in a systematic
way with bombarding energy as well. 

In summary, we applied the GLV parton energy loss~\cite{glv2,glv2b} in LO pQCD
to pion and kaon production in nuclear collisions and calculated the
sensitivity of the hadronic ratios to this property of the QCD plasma.  The
$K^+/\pi^+$ and $K^-/K^+$ ratios in the $2< p_T<10 $ GeV/c range 
were found to be the most sensitive to the gluon opacity at RHIC energies.

We want to compare our theoretical results to experimental data at RHIC
energies. Unfortunately, final data are available only 
at  $p_T< 2$ GeV/c in $Au+Au$ collisions at
$\sqrt{s} = 200$ AGeV~\cite{PHENIXpik}.
Looking into these data we find surprisingly large values for the
kaon to pion ratios at $p_T=2$ GeV/c in the most central 
collisions ($0-5 \  \%$),
namely 
$K^+/\pi^+ = 0.75 \pm 0.05$ and $K^-/\pi^-=0.65 \pm 0.05$.
Even in peripheral collisions ($60-92 \ \%$) a moderate value,
$0.45 \pm 0.07$ has been measured in both cases.
These values are larger then any of our calculated ratios
up to opacity $L/\lambda = 4$ in the low momentum region, see Figure 6. 
This indicates
that pion and kaon production is not dominated by fragmentation
in this transverse momentum region.
Furthermore, at low $p_T$ we see from the data
that $K^-/K^+ \approx 1$~\cite{PHENIXpik}. Figure 5
displays that this ratio should drop to a small value around $p_T=4 $ GeV/c
in case of jet fragmentation. Thus one can determine from
the drop of $K^-/K^+$ ratio the recovery of the fragmentation region.
It is interesting to note that 
${\overline p}/p$ ratio is constant ($\approx 0.75$)
up to $p_T \leq 6 $ GeV/c~\cite{PHENIXpik,PHENIXpap,STARpap}, 
which may indicate the lower limit of the 
fragmentation region. The measured anomalously high $p/\pi^+$ and
${\overline p}/\pi^-$ ratios in the intermediate momentum region
$2 < p_T < 6$ GeV/c at RHIC energies~\cite{PHENIXpap}
have been explained theoretically by quark coalescence/recombination 
models~\cite{HGF1,HGF2,HGF3}. Thus we may expect the application of
similar models for mesonic ratios in the same momentum region.
However, for $p_T > 6$ GeV/c our calculations with fragmentation
will be valid and applicable to extract the opacity
of the hot dense parton matter produced in Au+Au collisions.
\bigskip

Acknowledgments. We thank G. David and P. Stankus
for useful discussions on RHIC experiments.
This work was supported by the OTKA Grant No. T043455, 
T047050, NK062044,
the MTA-NSF-OTKA INT-0435701
and partly by the DOE Research Grants
U.S. DE-FG02-86ER-40251 and DE-FG02-93ER-40764.

\vfill\eject
\end{document}